\documentclass[twoside]{article}
\usepackage{epsfig.sty}
\newcommand{\beq}{\begin{eqnarray}}
\newcommand{\eeq}{\end{eqnarray}}
\begin{document}
\title{Review of Dark Matter}
\author{Leonard S. Kisslinger\\
Department of Physics, Carnegie Mellon University, Pittsburgh PA 15213 USA.\\
Debasish Das\\
Saha Institute of Nuclear Physics,1/AF, Bidhan Nagar, Kolkata 700064, INDIA.}
\date{}
\maketitle
\date{}
\noindent
PACS Indices:11.30.Er,14.60.Lm,13.15.+g
\vspace{3mm}


\begin{abstract}
  In this review of Dark Matter we review dark matter as sterile neutrinos,
  fermions, with their present and possibly future detection via neutrino
  Oscillations. We review the creation of Dark Matter via interactions with
  the Dark Energy (quintesence) field.  We also review bosons as dark matter,
  discussing a proposed search for dark photons. Since photons are vector
  bosons, if dark photons exist at least part of dark matter are vector bosons.
  Ongoing experimental detection of Dark Matter is reviewed.
\end{abstract}

\section{Introduction}

The most important experiments which have estimated the amount of Dark
Matter in the present universe are Cosmic Microwave Background Radiation
(CMBR) experiments, discussed in the section 2.

There have been a number of theoretical models for the creation of Dark
Matter, which is reviewed in section 3. It is almost certain that sterile
nuetrinos are part of Dark Matter.

Experiments detecting sterile nuetrinos via neutrino oscillation and
a theoretical study of neutrino oscillations with 3 active and 3 sterile
neutrinos with the present results are discussed in section 4. Also a recent
search for sub-Gev Dark Matter by the MiniBooNE-DM Collaboration is briefly
discussed in section 4.

Neutrinos are fermions with quantum spin 1/2. It is possible that some Dark
Matter particle are vector bosons with quantum spin 1, like the photon. A
possibe Dark Photon Search is discussed in section 5. Future experimental
detection of Dark Matter is reviewed in section 6.

\section{Cosmic Microwave Background Radiation (CMBR)}

There have been many CMBR experiments, such as Refs\cite{cobe96,acbar08,quad10,
wmap13} which have estimated the total density of the present universe,
Dark Matter density, dark energy density, etc.

With $\Omega$ the density of the Universe and $\Omega=1.0$ for a flat Universe,
results from CMBR observations are:
\beq
\label{CMBR-results}
\Omega &=& 1.0023^{+.0056}_{-.0054} \nonumber \\
{\rm Dark\;Energy\;Density\;(vacuum\;energy)}&\simeq& 0.73 \nonumber \\
{\rm Dark\;Matter\;Density}&\simeq& 0.23 \nonumber \\
{\rm Baryon (Normal Matter)\;Density} &\simeq& 0.04 \nonumber \\ 
{\rm Age\;of\;the\;Universe}&\simeq& 1.37 {\rm\;\;billion\;years} \; .
\eeq
Therefore about 23 \% of the Universe is Dark Matter.

About 73 \% of the universe is dark energy, which is anti-gravity. Dark
Energy (Quintessence) is anti-gravity and produced inflation at a very
early time, which is why we now have an almost homogeneous universe.

Only about 4 \% of the universe is normal matter.
\newpage
\section{Creation of Dark Matter}

There have been many studies of the origin of Dark Matter mass. In order
to explain why the universe is highly homongeneous Guth\cite{guth81} introduced
what is now called the dark energy or quintessence field to produce inflation.
Peebles and Ratra\cite{pr88} studied the quintessence field. Farrar and
Peebles\cite{fp04} intersecting Dark Matter and the Dark Energy (quintessence)
field.

A recent study estimated Dark Matter mass created via Dark Energy
interaction during the Cosmological Electroweak Phase Transition
 (EWPT)\cite{kc14} at a time $t_{EWPT} = 10^{-11}$ s.
Using the Dark Energy Lagrangian ${\cal L}^{DE}$
and Dark Mass-Dark Energy Lagrangian ${\cal L}^{DM-DE}$
\beq
\label{L}
  {\cal L}^{DE}&=& \frac{1}{2}\partial_\nu \Phi_q \partial^\nu \Phi_q
-V(\Phi_q) \nonumber \\
{\cal L}^{DM-DE}&=& g_D \bar{\psi}^{DM} \Phi_q \psi^{DM} \; ,
\eeq
with $\Phi_q$ the quintessence field and $V(\Phi_q)$ obtained from
Refs\cite{pr88,fp04}. Depending on the choice of parameters Ref\cite{kc14}
estimated Dark Matter mass $M_{DM}(EWPT)$ at the present time $t=t_{now}$  as
\beq
\label{DM-DE-EWPT}
M_{DM}(EWPT) &\simeq& {\rm few\;GeV\;\;to\;140\;GeV} \ .
\eeq
With this range of values Dark Matter particles might be detected in the
near future.

More recently a study estimated Dark Matter mass created via Dark Energy
interaction during the Cosmological Quantum Chromodynamics Phase Transition
(QCDPT)\cite{kd19} using the Lagrangians in Eq(\ref{L}) at a time $t_{QCDPT} =
 10^{-4}$ s, with the result at $t=t_{now}$
\beq
\label{DM-DE-QCDPT}
M_{DM}(QCDPT) &\simeq& {\rm 0.5\;\;to\;3.5\;TeV} \ ,
\eeq
which is more than an order of magnitude larger than $M_{DM}(EWPT)$ as
$t_{QCDPT}\gg t_{EWPT}$.

\section{Neutrino Oscillations and Sterile Neutrino Parameters}

Sterile neutrinos are a well-known source of Dark Matter.
For many years the MiniBooNE Collaboration has studied neutrino oscillations.
The MiniBooNE  Collaboration\cite{mini13} using $\bar{\nu}_\mu \rightarrow
\bar{\nu}_e$ oscillations detected a sterile neutrino $\nu_4$
and estimated the mass difference between this sterile neutrino and a standard
neutrino as
\beq
\label{sterilemass}
\Delta m^2&=& m_4^2-m_1^4 \simeq 0.06 {\rm (eV)}^2 \; ,
\eeq
which is too small for sterile neutrino $\nu_4$ to be a WIMP (Weakly Interacting
Massive Particle).

In a calculation with three active and three sterile neutrinos\cite{lsk16},
an extension of work with six neutrinos\cite{tg07},
$\mathcal{P}(\nu_\mu\rightarrow \nu_e)$, the transition probability for a
muon neutrino to oscillate to an electron neutrino, was estimated.

With the notation for the three active neutrinos $\nu_e=\nu_1,\nu_\mu=\nu_2,
\nu_\tau=\nu_3$ and the three sterile neutrinos are $\nu_4,\nu_5,\nu_6$ 
\beq
\label{Pue-1}
 \mathcal{P}(\nu_\mu \rightarrow\nu_e) &=& Re[\sum_{i=1}^{6}\sum_{j=1}^{6}
U_{1i}U^*_{1j}U^*_{2i}U_{2j} e^{-i(\delta m_{ij}^2/E)L}] \; ,
\eeq
where neutrinos with flavors $\nu_e,\nu_\mu,\nu_\tau$ and three sterile 
neutrinos, $\nu_{s_1},\nu_{s_2},\nu_{s_3}$ are related to neutrinos with 
definite mass by
\beq
\label{f-mrelation}
      \nu_f &=& U\nu_m \; ,
\eeq
where $U$ is a 6x6 matrix. In Ref\cite{lsk16} the sterile-active neutrino
mixing angles were calculated, which should be useful for future searches
for sterile neutrinos as Dark Matter particles.

Recently the MiniBooNE-DM Collaboration has carried out a search for sub-Gev
Dark Matter at the Fermilab\cite{mbde18}, but no excess from background
predictions was observed.

\section{Dark Photon Search}

There is a worldwide race in the search for Dark Photons.

A Dark Photon would
be a vector boson (quantum spin=1), while a neutrino is fermion (quantum
spin=1/2, so Dark Photons would be a very different kind of Dark Matter
particle than a sterile neutrino.

The Dark Photon Search is described in Ref\cite{p25}. The LANL(Los Alamos
National Laoratory), P-25, PHENIX team is planning to search
for Dark Photons at Fermilab. The Direct Search for Dark Photons at Fermilab
is reviewed in Ref\cite{kl16}, which can be found by clicking on Ref\cite{kl16}
in ``talks and presentations'' in Ref\cite{p25}. Since the detection of
Dark Photons is quite difficult the LANL-P-25 PHENIX team is carefully planning
for the experiment. In Ref\cite{p25}, using EmCal hardware, one finds a
description of the hardware by Hubert van Hecke. Kun Liu, a co-PI, will plan
for the P-25 PHENIX team at Fermilab.

\section{Experimental Detection of Dark Matter}

Since we do not know what dark matter is, we need a diverse pool of
instruments and approaches to detect it\cite{Liu:2017drf,Cooley:2014aya}.
Experiments LUX~\cite{Akerib:2013tjd},
PandaX-II\cite{Tan:2016zwf}, PICO\cite{Amole:2016pye,Amole:2017dex},
DAMA/LIBRA\cite{Bernabei:2010mq,Bernabei:2013xsa,Bernabei:2018yyw},
SuperCDMS\cite{Agnese:2017njq}, CRESST-III~\cite{Abdelhameed:2019hmk},
which use direct detection methods, are reviewed.

The {\bf LUX} (Large Underground Experiment)~\cite{Akerib:2013tjd}
is a dark matter experiment that aims to detect the  WIMPs
(Weakly Interacting Massive Particles), a dark matter candidate. The
experiment is carried out 1.5 km underground under SURF (Sanford
Underground Research Facility). The LUX detector is a
two-phase xenon time-projection chamber (TPC), containing 370 kg of
xenon. The liquid xenon contains an array of
photosensors which can sense a single photon of light.

The second phase of the {\bf PandaX} (Particle and astrophysical Xenon)
project~\cite{Tan:2016zwf} is located in the China Jinping Underground
Laboratory (CJPL) and started it's operation in
early 2016. PandaX-II comprises of a 580 kg dual-phase xenon TPC along
with a 60x60 cm cylindrical target containing 55 top and 55 bottom
Hamamatsu R11410-20 3-inch photomultiplier tubes (PMTs).  The data sets
(Run 9 and Run 10)  are now used to perform
systematic studies using the PandaX-II detector. 

The {\bf PICO} project\cite{Amole:2016pye} is working towards the detection of
dark matter particles with the bubble chamber technique. The expriments are
installed at a depth of 2 km in the SNOLAB underground laboratory at Sudbury,
Ontario, Canada. Searching WIMPs using superheated bubble chambers which are
operated in thermodynamic conditions at which they are almost insensitive to
gamma or beta radiation is the main detection principle of PICO experiment.
A bigger version of the experiment with up to 500 kg of active mass is in
progress.

The DAMA experiment at the Gran Sasso National Laboratories of the I.N.F.N. is
an observatory for rare processes that use radiopure scintillator set-ups.
The main activity field of {\bf DAMA/LIBRA}\cite{Bernabei:2010mq,
  Bernabei:2013xsa,Bernabei:2018yyw} is the investigation on Dark Matter
particles in the galactic halo.  The second generation DAMA/LIBRA set-up
(Large sodium Iodide Bulk for RAre processes is continuing
the investigations of DAMA/NaI.

The {\bf SuperCDMS}(SCDMS)\cite{Agnese:2017njq} is the
next generation of the CDMS(Cryogenic Dark Matter Search) II experiment,
which was located underground in the deep Soudan mine of Minnesota, USA.
SCDMS wants to
re-locate at SNOLAB (Vale Inco Mine, Sudbury, Canada) which is a much deeper
facility. CDMS-II comprises of a cryostat surrounded by a passive shield and
an outer muon veto situated beneath an overburden of 2090 meters water
equivalent. The use of underground facilities provide the required shielding
from the cosmic events and hence reduce the interference of known background
particles. 

The {\bf CRESST}(Cryogenic Rare Event Search with Superconducting Thermometers)
experiment\cite{Abdelhameed:2019hmk}, located at the Gran Sasso underground
laboratory in Italy, has operated 13 detector modules in CRESST-III Phase 1
(2016-2018). CRESST searches for dark matter particles
via their elastic scattering off the nuclei in a target material.
The CRESST target comprises of scintillating $CaWO_{4}$ crystals.This
experiment searches for dark matter masses below 1.7 $GeV/c^2$, extending
the CRESST-II result in 2015. For CRESST-III, whose Phase 1 started in July
2016, detectors have been optomized to further probe the low-mass region.

The {\bf DINO} Dark Matter experiment\cite{dino16} will search for WIMPS
using scintillation detectors for the detection of recoiling nuclei.

The presence of low mass WIMPs in the various ongoing
experiments~\cite{lowmass1} have generated motivations to
cross-check future proposals\cite{Seth:2019sff}.
\newpage

\section{Conclusions}

In this Review we have reviewed many aspects of experiments and theory related
to Dark Matter.

Cosmic Microwave Background Radiation (CMBR) experiments have shown that about
23 \% of the Universe is Dark Matter and only about 4 \% of the universe is
normal matter.

The creation of Dark Matter mass during the EWPT and QCDPT, two Cosmological
Phase Transistions, via the dark matter field interacting with the dark
enargy (quintessence) field are reviewed. The Dark Matter mass $M_{DM}$ created
during the QCDPT could be more than 1 TeV,  more than an order of magnitude
larger than $M_{DM}$ created during the EWPT.

Theoretical and experimental studies of neutrino oscillations producing
sterile neutrinos are reviewed, with massive sterile neutrinos possibly
being the Dark Matter detected by CMBR experiments. Also a Dark Photon search,
with photons vector bosons rather than sterile neutrinos, which are fermions,
is briefly reviewed.

Finally, six experiments for the experimental detection of Dark Matter are
reviewed, with a detailed explanation of how these experiments can detect
Dark Matter.

\vspace{5mm}
\Large
{\bf Acknowledgements
\normalsize
\vspace{5mm}

Author D.D. acknowledges the facilities of Saha Institute of Nuclear Physics, 
Kolkata, India. Author L.S.K. acknowledges support in part as a visitor at Los
Alamos National Laboratory, Group P25.

\end{document}